\documentclass[journal=jacsat,manuscript=article]{achemso}
\usepackage[version=3]{mhchem} 
\usepackage{xcolor}
\usepackage{soul} 

\usepackage{graphicx}%
\usepackage{multirow}%
\usepackage{amsmath,amssymb,amsfonts}%
\usepackage{amsthm}%
\usepackage{mathrsfs}%
\usepackage[title]{appendix}%
\usepackage{xcolor}%
\usepackage{textcomp}%
\usepackage{manyfoot}%
\usepackage{booktabs}%
\usepackage{algorithm}%
\usepackage{algorithmicx}%
\usepackage{algpseudocode}%
\usepackage{listings}%

\title{A Theoretical Framework for the Coupling of Macroscale-Nanoscale Mechanochemical Phenomena in Condensed Matter}
\author{Brenden W. Hamilton}
\affiliation[Los Alamos National Laboratory]
{Theoretical Division, Los Alamos National Laboratory, Los Alamos, New Mexico 87545, USA}
\email{brenden@lanl.gov}

\date{\today}

\begin{document}


\singlespacing
\begin{abstract}

The field of covalent mechanochemistry has transitioned from fundamental science to engineering applications, yet it lacks a robust theoretical framework for predicting reaction kinetics in condensed matter. 
Existing analytical models fail under realistic conditions where macroscopic strains drive molecular-scale deformations that are highly non-linear. 
We develop a non-perturbative theoretical framework that captures activation barrier changes in highly strained molecules undergoing complex, non-linear deformations, describing the macroscale-nanoscale coupling of phenomena. 
The framework yields general expressions, parameterizable from atomistic simulations, enabling multiscale prediction of mechanochemical behavior.
By presenting the expressions in terms of general observables, this work enables predictions of mechanochemical effects from simple structure optimization calculations, enabling the use of high level quantum chemical methods.
We demonstrate this approach on the mechanochromic polymer spiropyran, showing how non-linear strain fields govern mechanophore activation.

\end{abstract}

\newpage

Mechanochemistry, broadly, is the application of a force in order to influence a chemical reaction.
This non-thermal alteration to reactions have been widely utilized as both the sole driver of the reaction and as accompanying a thermal driver for chemical reaction\cite{Ribas-Arino2012Review,james2013mechanochemistry}. 
Mechanochemistry can include a wide range of methods such as ball milling to alter structures\cite{speight2025ball,frivsvcic2020mechanochemistry},
sonication to induce strains\cite{cravotto2013mechanochemical,hwang2025engineering}, and atomic force microscopy to strain individual bonds\cite{beyer2005mechanochemistry,wang2014remote}.

The sub-field of covalent mechanochemistry is the space in which direct straining of chemical bonds leads to an alteration in a chemical reaction, typically accelerating it.
The mechanical extension of a covalent bond leads to changes in the local electronic structure which can flatten reaction barriers\cite{Stauch2016Review,o2021many,howard2018switching,barbee2018substituent,hamilton2023interplay,sun2024charge}
and make available previously suppressed\cite{Hamilton2022ManyBody,Hamilton2022Extemporaneous} or forbidden reactions\cite{Wang2015Inducing,Ong2009FirstPrinciples}.
The seminal work "How Strong is a Covalent Bond"\cite{Grandbois1999HowStrong} greatly expanded this field by directly straining a single bond until scission in an atomic force microscope.

Over the last two decades, covalent mechanochemistry has moved from a fundamental science to an applied one, with applications in functional polymers\cite{Deneke2020Engineers,kryger2011structure,lee2015relative,chu2019polysilsesquioxane,pala2008nonvolatile}, biological systems\cite{yum2009mechanochemical,chen2025ph},
sustainable engineering\cite{Ardila-Fierro2021Sustainability,arfelis2024linking,chen2026mechanochemically}, drug-delivery systems\cite{shi2024polymer,rehfeldt2007cell,luo2025unidirectional},
and catalysis for manufacturing\cite{ralphs2013application,buyanov2009mechanochemical}.
A large majority of these processes are driven by macroscopic strains applied to condensed matter to drive molecular strains at the nanoscale.

Simple models such as that from Bell\cite{Bell1978Models} have equated the change in barrier height to the work done by the applied force along the reaction coordinate:
$$\Delta E^{\ddagger}_F = \Delta E^{\ddagger}_o - W_\xi = \Delta E^{\ddagger}_o - F \cdot \Delta \xi $$
where the applied force, $F$, is assumed to be \textit{exactly} along the reaction coordinate $\xi$,
$\Delta E^{\ddagger}_o$ is the reaction barrier without an applied force and $ W_\xi$ is the work done on the system by the applied force.
The tilted PES model\cite{Ribas-Arino2009Understanding} allows for a change in stationary states, but still assumes $F$ is along $\xi$.
A mounting body of work has begun to show that, especially in condensed matter, the deformations to molecules can be complex and multi-dimensional, described as a combination of bending and torsion degrees of freedom, where only a subset of the forces will be related to chemical 
reactions and will not inherently be pure pulling forces on a scissile 
bond\cite{Kroonblawd2020ShearBands,Hamilton2021HotspotsBetterHalf,hamilton2022rapid,centellas2024mechanochemically,Kingsbury2011Shear,Silberstein2013Modeling,Sung2018Interfacial,Larsen2013FlexActivated,huang2026mechanochemical,janissen2022mechanochemistry,hamilton2023intergranular,hamilton2023influence,hamilton2023using}.
Hence, the key knowledge gap here is that the assumption of the applied force being along the reaction coordinate $\xi$ is typically not valid for engineering materials.
The primary issue of these engineering conditions for mechanochemistry is that there lacks a more robust theoretical and analytical framework
for the prediction of reaction kinetics due to macroscopic forces being transduced onto covalent bonds.

Here we develop a non-perturbative, theoretical framework for modeling the changes in activation barrier for reactions in highly strained molecules due to complex, non-linear deformations.
The change in barrier is expressed as due to the rise in the molecular strain energy, the result of the macroscopic strain response of the overall system, $\frac{\partial \Delta E^{\ddagger}_o}{\partial \Delta U_{sys}}$.
This can be expanded as a product of more readily available observables, which fundamentally reduces the complexity of the problem,
enabling more direct modeling of deformations that are not directly along the reaction coordinate.
Here, the reaction barrier change is expressed as either $\frac{\partial \Delta E^{\ddagger}_o}{\partial F_{\xi}} \frac{\partial F_\xi}{\partial \Delta U_{sys}}$
or $\frac{\partial \Delta E^{\ddagger}_o}{\partial \xi} \frac{\partial \xi}{\partial \Delta U_{sys}}$.
The former is readily applicable when collective variables can be readily designed such that the resultant force along the reaction path is explicitly representable, typically requiring force analysis tools\cite{Stauch2016Review,Stauch2014JEDI,Stauch2016Knots}.
The latter is more readily practical for simulation workflows and will be shown to greatly lower the total cost of simulations needed, enabling the use of this route for expensive quantum chemistry methods. 


To derive these quantities, one can define a set of molecular scale variables, where $\epsilon_{sys}$ is the total strain on the molecule and $\Delta U_{sys}$ is the total internal energy rise of the molecule.
$\Delta E^{\ddagger}_{eff}$ is the effective activation energy of the reaction, as altered by the strain to the molecule.
We recast to this strain energy formalism instead of the standard applied force following the activation-strain model\cite{bickelhaupt2017analyzing}, 
in which the activation barrier can be calculated as the sum of the energetic increases from the required strain to reach the transition state and altered interactions of the strained system.
The strain energy applied here works in the same sense, driving the system towards the strained TS configuration.
Depending on how the strain is applied to the molecule, there will be some change to the remaining or "effective" activation barrier.
If the strain has no connection with the intrinsic reaction coordinate, it should not alter the reaction at all.

We define this effective activation barrier as:
$$\Delta E^{\ddagger}_{eff} = \Delta E^{\ddagger}_o - \frac{\partial \Delta E^{\ddagger}_o}{\partial \Delta U_{sys}} \Delta U_{sys}$$
The complexity of the problem arises from attempting to define $\frac{\partial \Delta E^{\ddagger}_o}{\partial \Delta U_{sys}}$,
which could be done by arbitrarily fitting to experiments or molecular dynamics simulations.
However, as $U_{sys}$ is a function of $\epsilon$, resulting in a different $\frac{\partial \Delta E^{\ddagger}_o}{\partial U_{sys}}$ for each possible $\epsilon$, and this quickly becomes an onerous task.
By defining $F_\xi$ as the force applied along the implicit reaction coordinate $\xi$, we can use the chain rule to expand
$$\frac{\partial \Delta E^{\ddagger}_o}{\partial \Delta U_{sys}} = \frac{\partial \Delta E^{\ddagger}_o}{\partial F_{\xi}} \frac{\partial F_\xi}{\partial \Delta U_{sys}}$$
where we can define $\frac{\partial \Delta E^{\ddagger}_o}{\partial F_{\xi}}$ as any one of the simple expressions existing in the literature.
For simplicity, we use the model from Bell\cite{Bell1978Models}, giving us:
$$\frac{\Delta E^{\ddagger}_o}{\Delta F_{\xi}} =  \Delta \xi$$
leaving us only to define $\frac{\partial F_\xi}{\partial \Delta U_{sys}}$.

By defining the $F_{sys}$ as the total forces on the system from $\Delta U_{sys}$ and $F_\xi$ and the projection of that force along the reaction coordinate $\xi$ we end up with :
$$\frac{\partial F_\xi}{\partial \Delta U_{sys}} = s * \frac{|\text{proj}_\xi(\epsilon)|^2}{|\epsilon|^2 |\xi|}$$
resulting in the full terms as:
$$\frac{\partial \Delta E^{\ddagger}_o}{\partial \Delta U_{sys}} = s * \frac{|\text{proj}_\xi(\epsilon)|^2}{|\epsilon|^2}$$
where $s$ is the sign of projection, which is $+1$ or $-1$ depending if the force along $\xi$ is parallel or anti-parallel to $\xi$, respectively.
The full derivation is available in the Supplemental Material.
Interestingly, in the case where both $\epsilon$ and $\xi$ can be described as single cartesian vectors, where $\theta$ is the angle between them, the entire term reduces to: $s * cos^2(\theta)$

However, in most cases, explicitly or analytically defining this projection for anything more complex than a single bond extension becomes numerically difficult.
The alternative approach is to break up the $ \frac{\partial \Delta E^{\ddagger}_o}{\partial \Delta U_{sys}}$ term as
$$\frac{\partial \Delta E^{\ddagger}_o}{\partial \Delta U_{sys}} = \frac{\partial \Delta E^{\ddagger}_o}{\partial \xi} \frac{\partial \xi}{\partial \Delta U_{sys}}$$
Where, for a reasonable choice of the implicit reaction coordinate, both terms are easily extractable from simple molecular dynamics simulations.
Additionally, whereas $\frac{\Delta E^{\ddagger}_o}{\Delta F_{\xi}}$ is different for each deformation, $\frac{\partial \Delta E^{\ddagger}_o}{\partial \xi}$ is independent of $\epsilon$,
meaning one must sample this for only one deformation, such as where a simple force is applied directly along $\xi$.
Then $\frac{\partial \xi}{\partial \Delta U_{sys}}$ is trivial to extract from steered MD simulations.

Overall, the $\frac{\partial \Delta E^{\ddagger}_o}{\partial F_{\xi}} \frac{\partial F_\xi}{\partial \Delta U_{sys}}$ form proves more practical in cases where $\frac{P(\epsilon, \xi)}{|\epsilon| |\xi|}$ can be readily defined. 
These would typically be general cases in which $\xi$ and other collective variables can be well modeled within simple systems following the harmonic approximation or other definitions in which representations of the molecular system
make $\frac{\partial F_\xi}{\partial \Delta U_{sys}}$ easy to define.
The $\frac{\partial \Delta E^{\ddagger}_o}{\partial \xi} \frac{\partial \xi}{\partial \Delta U_{sys}}$ form is more directly applicable to simulation workflows,
as it relies on more readily available observables.
This can be directly applied to quantum chemical calculations in which $\xi$ is known the collective variable can be explicitly and simply mapped in a cartesian representation.
We will spend the rest of this article exploring the application of the latter to the mechanophore spiropyran (SP), in which the molecule is pulled along its backbone while an additional torsional deformation is applied,
resulting in a complex strain $\epsilon$ where only a part of the torsional deformation projects onto $\xi$.
Spiropyran is a key example, as it is a mechano-chromic molecule that is employed in condensed matter and readily used for engineering purposes\cite{Deneke2020Engineers,Davis2009ForceInduced}, but is often computationally studied as an isolated molecule.


For molecular dynamics simulations of SP, a total of 5 different molecular deformation paths are chosen to add to the pulling of the molecule.
Figure 1 displays the 5 deformations, with atoms involved colored in purple that make up a dihedral that is perturbed with an external field prior to, and during, the linear deformation applied along the backbone via the COGEF method\cite{Grandbois1999HowStrong,Beyer2000Mechanical}.
For the deformation named 'Central Twist", the dihedral causes a rotation around the spiro atom, rotating the left two rings with respect to the plane of the right two.
For the "Secondary Ring Buckle", the rightmost ring structure is buckled out of plane.
For all others, the left two and right two rings are folded towards or away from each other with the spiro atom acting as a hinge,
with the "Central Buckle" pushing the two halves to more of a right angle in the plane of the image and the two "Penta" ring based buckle deformations drive the rings into the page and rightwards towards to oxygen atom bonded to the spiro atom.
Each of these is chosen to represent some sort of deformation path that could plausible occur in condensed matter conditions, with previous molecular dynamics simulations supporting these choices\cite{hamilton2022rapid}.

\begin{figure}[htpb]
 \includegraphics[width=0.8\textwidth]{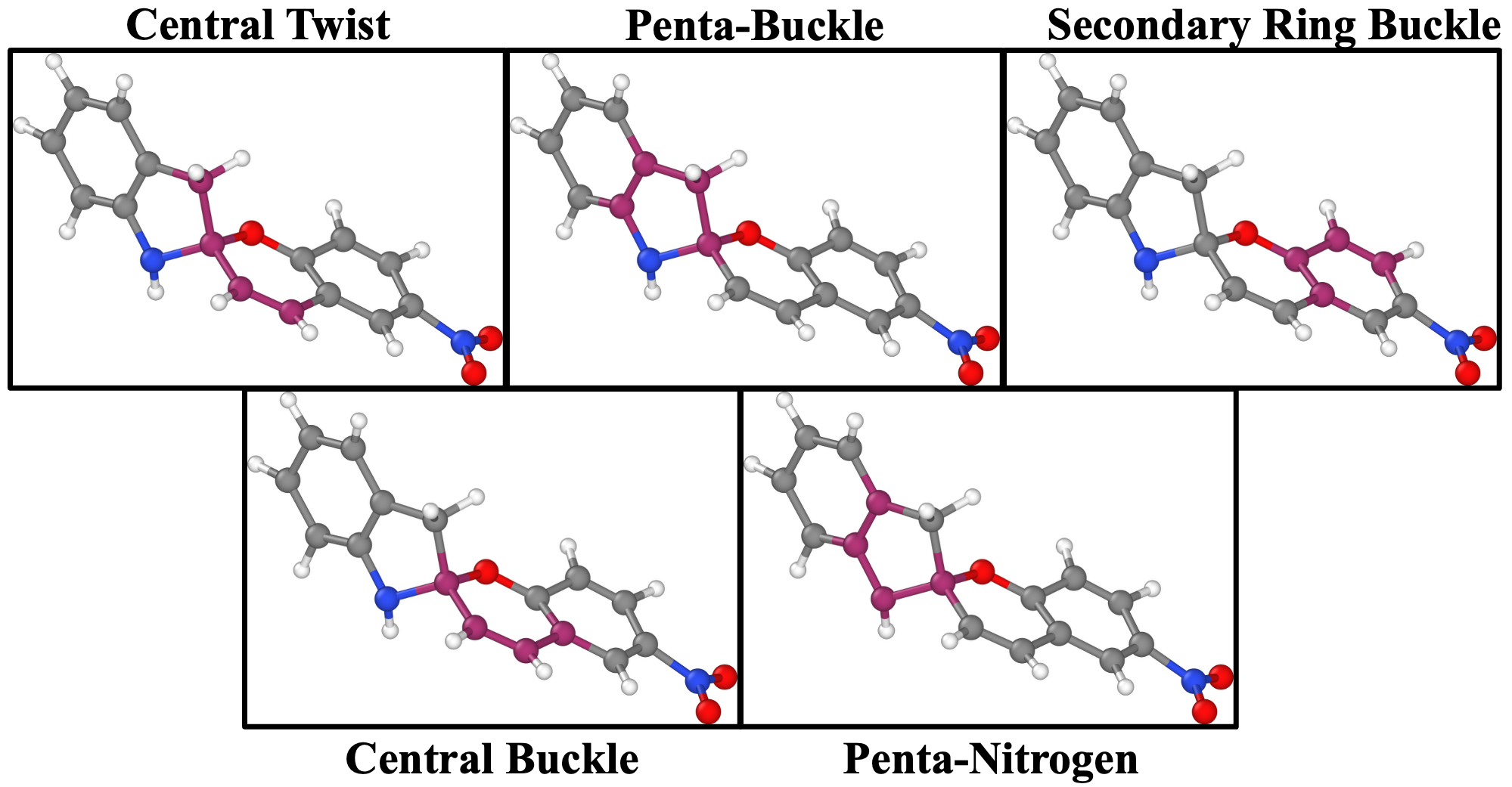}
   \caption{Atoms in the spiropyran molecule involved in the torsional deformation for each deformation path.}
 \label{fig:Fg5}
\end{figure}

All simulations are performed using classical molecular dynamics with the LAMMPS simulation package\cite{Thompson2022LAMMPS} and the ReaxFF\cite{VanDuin2001ReaxFF,Liu2011ReaxFFLG} forcefield, which has been previously used for spiropyran pulling reactions\cite{Hamilton2022ManyBody}.
To induce the reaction via pulling, the CoGEF (Constrained Geometries to Simulate External Force) method was used\cite{Beyer2000Mechanical,Grandbois1999HowStrong}.
To induce the additional deformations not along the reaction path, the Many-Bodied Steered Molecular Dynamics (MBsMD) approach was used\cite{Hamilton2022ManyBody,hamilton2023interplay}.
Extended details of methods are available in the Supplemental Materials.

For each case, the size of the deformation is modulated by the strength of the external field acting on the torsional motion.
The torsional deformation is then equilibrated and the sample deformed along the backbone axis until scission of the C-O bond with the spiro atom, which is the reaction that occurs in all cases.

Figure 2 shows, for each of the deformations, how the reaction barrier changes with increasing deformation energy rises.
The x-axis is the system relaxed energy rise, i.e. how much higher in energy the optimized
structure with the external field is than the baseline optimized structure.
This directly represents $\frac{\partial \Delta E^{\ddagger}_o}{\partial \Delta U_{sys}}$ directly sampled from MD.

The five deformations group into three clusters of response, a monotonic lowering of the barrier, a monotonic
increasing of the barrier, and little to no effect.
As somewhat expected from the design, the "Secondary Ring Buckle" is the deformation that has no effect.
The presumptive implicit reaction coordinate is the stretching of the C-O bond and the ring opening paths, and
per the equations derived above, as this deformation does not project much onto the reaction its effect on the reaction is minimal.

From Figure 2, we show the Central Twist and Central Buckle deformations both lower the activation barrier.
They both give a mostly linear response with similar slopes, showing strong correlation between the deformation path and the reaction path.
Both deformations applied to the 5 carbon ring section of the spiropyran molecule result in an increase in the activation barrier.
These increases are much smaller in magnitude than the decreasing deformations and give a slightly nonlinear response for small deformations.

\begin{figure}[htpb]
 \includegraphics[width=0.4\textwidth]{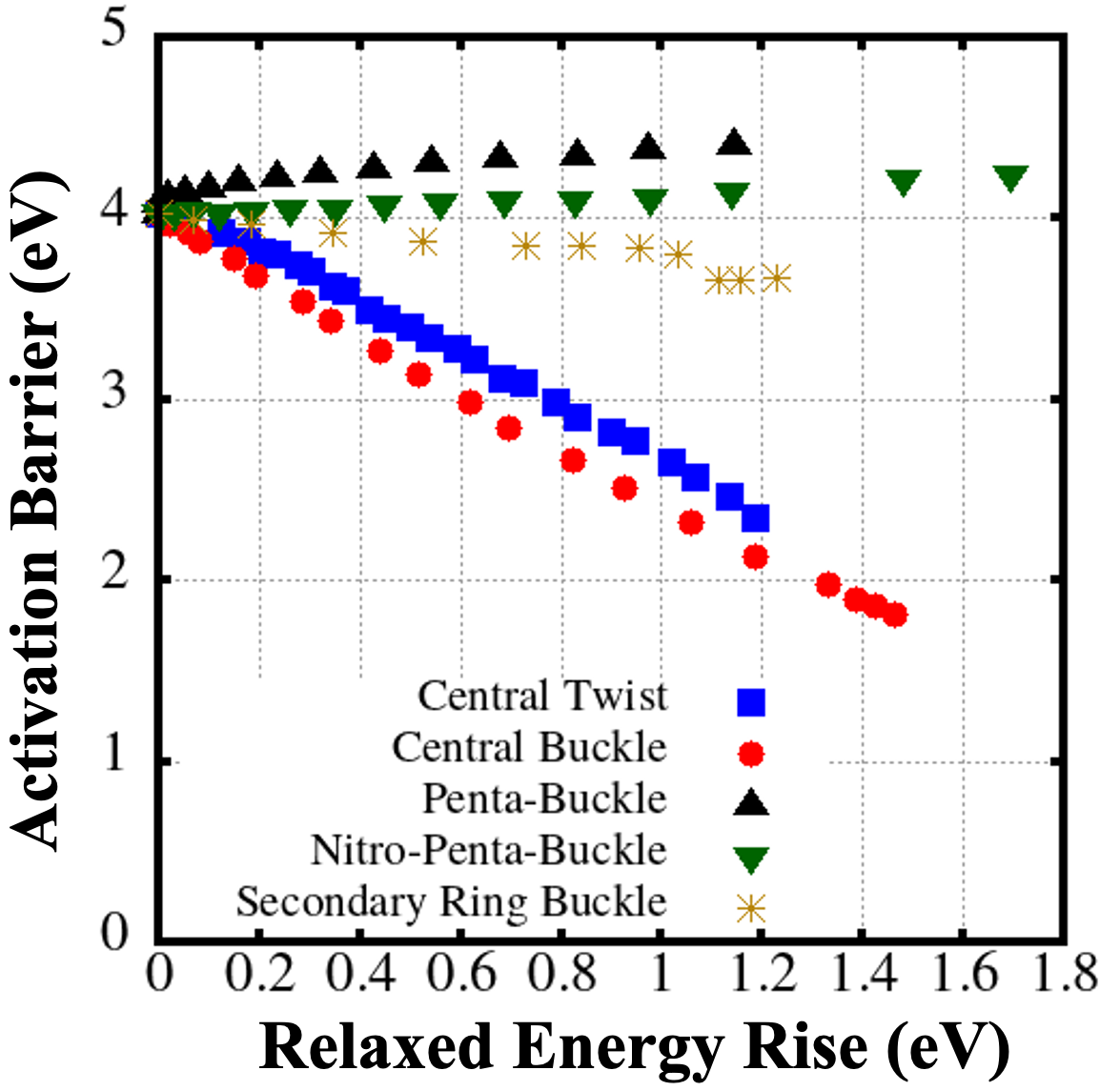}
   \caption{The effective activation barrier of the ring opening reaction for each deformation for a given internal energy rise from the deformation.}
 \label{fig:Fg2}
\end{figure}

To better assess how these deformations interact with the implicit reaction coordinate, we chose a CV of the C-O bond distance for simplicity.
Figure 3 represents an analytical approach to assessing $\frac{\partial \Delta E^{\ddagger}_o}{\partial \xi} \frac{\partial \xi}{\partial \Delta U_{sys}}$, breaking the data from Figure 2 into the two partial derivatives.
The left panel, representing $\frac{\partial \Delta E^{\ddagger}_o}{\partial \xi}$, cleanly shows all 5 deformation paths falling onto a single line.
This validates the assumption above that $\frac{\partial \Delta E^{\ddagger}_o}{\partial \xi}$ is independent of $\epsilon$and $U$.
Small deviations from the line, especially for the "Central Twist" deformation at large deformations, can most likely be attributed to $\xi$ and the C-O bond distance not being exactly the same.
The key importance of this result is that $\frac{\partial \Delta E^{\ddagger}_o}{\partial \xi}$ can be readily approximated by a simple linear function, requiring only a valid description of $\frac{\partial \xi}{\partial \Delta U_{sys}}$,
which can be calculated by constrained geometry optimization.
This abates the need for computationally expensive molecular dynamics simulations and fundamentally reduces the complexity of the problem.

The right panel of Figure 3 represents $\frac{\partial \xi}{\partial \Delta U_{sys}}$, providing the analytical mapping between the large scale strain applied to the whole molecule
Similar to Figure 2, there is a strong difference in the magnitude and trends of each deformation, showing how the deformation projects onto the reaction coordinate.
For the "Secondary Ring Buckle", the chosen reaction coordinate does not change with increasing molecular strain, which is what leads to no change in the reaction barrier.
For both 5 carbon rig deformations, increasing the deformation compresses the C-O bond, directly leading to the rise in activation energy for breaking the bond.
For the two deformations that lower the barrier significantly, there is a start difference in the trend with increasing strain/energy.
The "Central Twist" deformation results in super-linear growth, showing that the coupling between the deformation path the reaction coordinate is strong and that it is a highly efficient mechanism for inducing the reaction.
The "Central Buckle" deformation shows a sub-linear response, providing diminishing returns with continued deformation.
The stems from a variety of molecular relaxation mechanisms where buckling of the ring can lead to structural relaxations that suppress additional C-O bond stretch.

\begin{figure}[htpb]
 \includegraphics[width=0.8\textwidth]{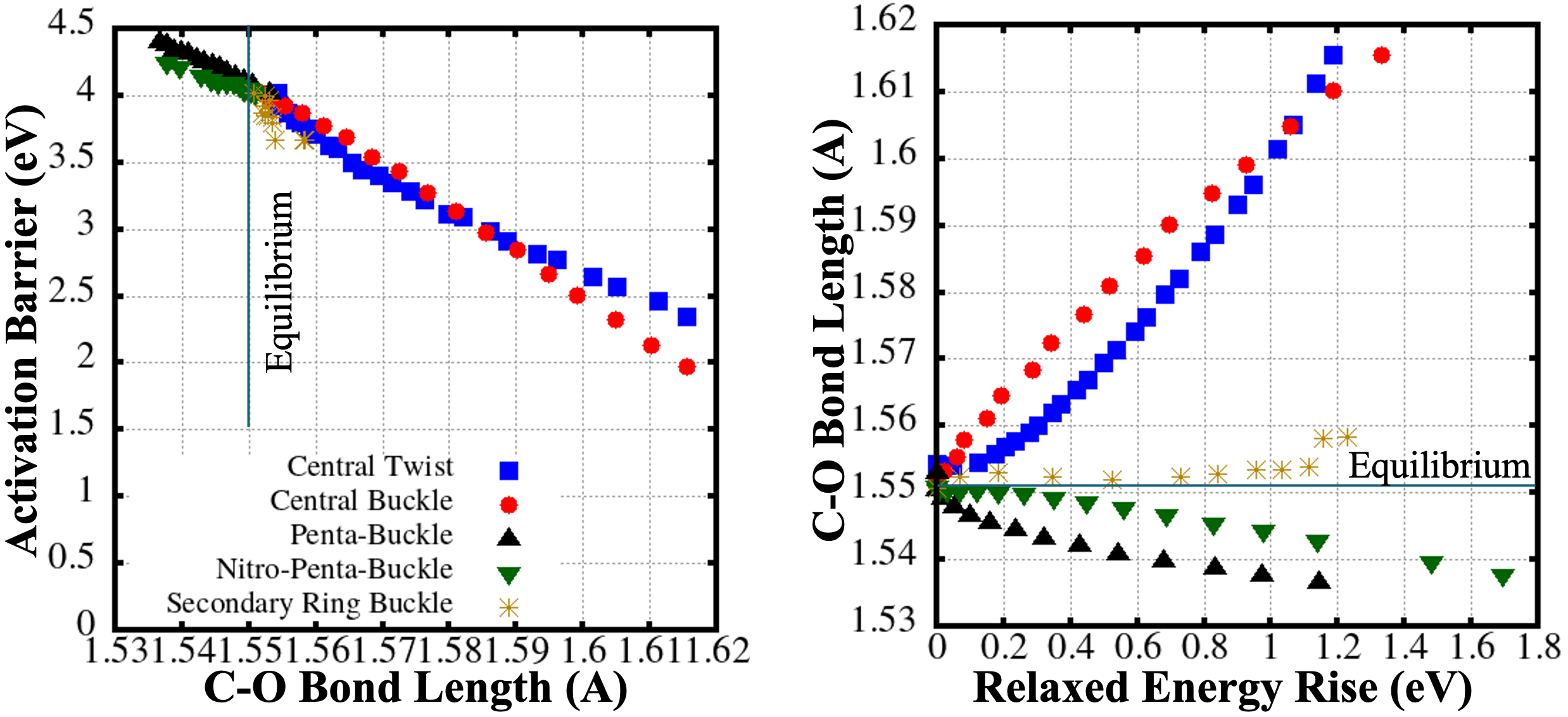}
   \caption{The effective activation barrier as a function of the reaction coordinate and the reaction coordinate as a function of the internal energy rise from the deformation, showing the chain rule version of Figure 2.}
 \label{fig:Fg3}
\end{figure}

Overall, here, we show that through a simple chain rule expansion, one can derive expressions that describe how complex, intra-molecular strains can influence chemical reactions via mechanochemical transduction of forces onto the reaction path.
The key advance here is that, while previous analytical expressions have the limiting assumption that the applied force $F$ is directly along the reaction coordinate $\xi$, the framework presented here
can be used for any direction and level of complexity of a molecular deformation $\epsilon$.
Using the technique described here, the changes in reactivity in a molecular system as a function of molecular strain can be expanded into a product of more readily available observables,
enabling more analytical, interpretable, and accessible predictions of reaction chemistry.
Molecular dynamics simulations validate the assumptions made here to derive the functional forms and show that separating out the
strain energy rise and the activation barrier change, via a chain rule, using the change in an implicit reaction coordinate can reduce the complexity of the problem by introducing an extra variable.
By expanding $\frac{\partial \Delta E^{\ddagger}_o}{\partial \Delta U_{sys}}$ into $\frac{\partial \Delta E^{\ddagger}_o}{\partial \xi} \frac{\partial \xi}{\partial \Delta U_{sys}}$, we show that $\frac{\partial \Delta E^{\ddagger}_o}{\partial \xi}$ can be approximated as a simple linear
response to $\frac{\partial \xi}{\partial \Delta U_{sys}}$, which can be mapped by assessing structural relaxations, removing the need to model the full reaction for each deformation path and magnitude.

While this is broad utility in this method for modeling reactions in strained molecules, significant future work is needed to assess things such as the statistical variance and fluctuations that come from
applying these ideas to finite temperature domains and in large scale condensed matter where each strained molecular will be in a different state, resulting in modeling distributions of different intra-molecular strain states
within some representative volume element.

\section{Acknowledgments}

The author thanks Sergei Tretiak and Travis Jones for insightful discussions on the topics discussed here.
Funding for this project was provided by the Director’s Postdoctoral Fellowship program at Los Alamos National Laboratory, project LDRD 20220705PRD1. 
Partial funding was provided by the Advanced Simulation and Computing Physics and Engineering Models project (ASC-PEM). This research used resources provided by the Los Alamos National Laboratory (LANL) Institutional Computing Program. 
This work was supported by the U.S. Department of Energy (DOE) through LANL, which is operated by Triad National Security, LLC, for the National Nuclear Security Administration of the U.S. Department of Energy (Contract No. 89233218CNA000001). 
Approved for Unlimited Release LA-UR-26-26166.

\includegraphics[scale=0.9, trim=0.75in 0.75in 0.75in 0.0in, clip, page=1]{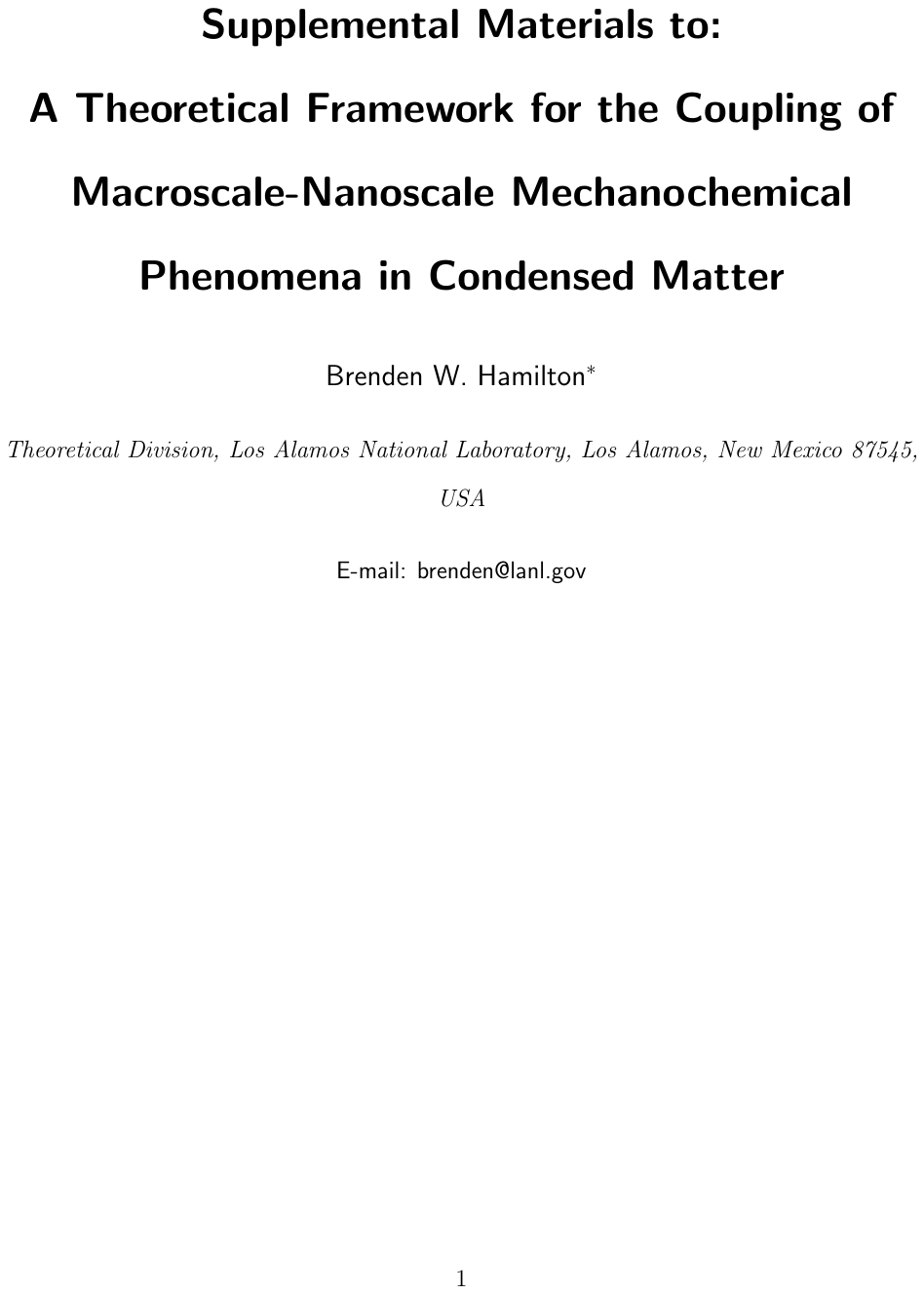}\\
\includegraphics[scale=0.9, trim=0.75in 0.75in 0.75in 0.0in, clip, page=2]{Supplemental_Materials.pdf}\\
\includegraphics[scale=0.9, trim=0.75in 0.75in 0.75in 0.0in, clip, page=3]{Supplemental_Materials.pdf}\\
\includegraphics[scale=0.9, trim=0.75in 0.75in 0.75in 0.0in, clip, page=4]{Supplemental_Materials.pdf}\\
\includegraphics[scale=0.9, trim=0.75in 0.75in 0.75in 0.0in, clip, page=5]{Supplemental_Materials.pdf}\\
\includegraphics[scale=0.9, trim=0.75in 0.75in 0.75in 0.0in, clip, page=6]{Supplemental_Materials.pdf}\\
\includegraphics[scale=0.9, trim=0.75in 0.75in 0.75in 0.0in, clip, page=7]{Supplemental_Materials.pdf}\\

\bibliography{references}

\end{document}